\documentclass[
 reprint, 
superscriptaddress,
longbibliography,
 amsmath,amssymb,
 aps, physrev,
floatfix,
]{revtex4-2}

\usepackage{graphicx}
\usepackage{dcolumn}
\usepackage{bm}
\usepackage[colorlinks=true,linkcolor=blue,citecolor=blue,urlcolor=blue]{hyperref}

\begin{document}

\preprint{APS/123-QED}

\title{\textbf{Electric-field-induced X-ray Nonreciprocal Dichroism in Hematite} 
}%

\author{Takeshi Hayashida}
\thanks{T. Hayashida and K. Matsumoto contributed equally to this work.\newline
Corresponding author: takeshi.hayashida@ru.nl}
\affiliation{HFML-FELIX, Radboud University, Toernooiveld 7, 6525 ED Nijmegen, the Netherlands}
\affiliation{Department of Applied Physics, University of Tokyo, Bunkyo-ku, Tokyo 113-8656, Japan}

\author{Koei Matsumoto}
\affiliation{Department of Applied Physics, University of Tokyo, Bunkyo-ku, Tokyo 113-8656, Japan}

\author{Keito Arakawa}
\affiliation{Department of Advanced Materials Science, University of Tokyo, Kashiwa, Chiba 277-0882, Japan}

\author{Yves Joly}
\affiliation{Université Grenoble Alpes, CNRS, Institut Néel, F-38042 Grenoble, France}

\author{Sergio~Di\ Matteo}
\affiliation{Univ Rennes, CNRS, IPR, UMR 6251, F-35000 Rennes, France}

\author{Kenji Tamasaku}
\affiliation{RIKEN SPring-8 Center, Sayo, Hyogo 679-5148, Japan}

\author{Yoshikazu Tanaka}
\affiliation{RIKEN SPring-8 Center, Sayo, Hyogo 679-5148, Japan}

\author{Tsuyoshi Kimura}
\affiliation{Department of Applied Physics, University of Tokyo, Bunkyo-ku, Tokyo 113-8656, Japan}

\date{\today}

\begin{abstract}
Hematite ($\alpha$-$\mathrm{Fe_2O_3}$) is a prototypical room-temperature antiferromagnet whose time-reversal-odd magnetic structure has recently attracted renewed attention. While such magnetic symmetry can be characterized in terms of higher-order multipoles beyond the magnetic dipole, their manifestation in measurable physical phenomena has remained largely elusive. In this work, we investigate x-ray absorption near the Fe \textit{K}-edge of hematite under an applied electric field, which explicitly breaks space-inversion symmetry. We observe an electric-field-induced x-ray nonreciprocal linear dichroism (\textit{E}-induced XNLD) that reflects the time-reversal-odd nature of the magnetic order. Numerical simulations based on \textit{ab-initio} density functional theory reproduce the observed spectra, including their dependence on the antiferromagnetic domain and x-ray polarization. Furthermore, a symmetry-resolved multipole analysis reveals that this response originates from the magnetic quadrupole and the magnetic toroidal octupole induced by the applied electric field. These results demonstrate that electric-field-modulated x-ray absorption provides direct access to the antiferroic order of higher-order multipoles in time-reversal-odd antiferromagnets, thereby establishing a general framework to uncover hidden symmetry properties in magnetic materials. 
\end{abstract}

\maketitle


\section{\label{sec:level1}INTRODUCTION}

Hematite ($\alpha$-$\mathrm{Fe_2O_3}$) is a ubiquitous, yet prototypical antiferromagnetic (AFM) material that has been studied for decades. In earlier studies, its weak ferromagnetism originating from the Dzyaloshinskii-Moriya (DM) interaction \cite{Neel1948-uk, Neel1949-ox, Dzyaloshinsky1958-eh, Dzyaloshinsky1958-zj, Moriya1960-dq} and the Morin transition \cite{Morin1950-dl} were extensively investigated \cite{Morish1994-nw}. More recently, various distinct physical properties and functionalities have been reported, including electric switching of magnetization \cite{Cheng2020-cc, Cogulu2021-lq}, anomalous Hall electrical transport  \cite{Galindez-Ruales2025-ja}, magneto-optical Kerr effect \cite{Pan2026-gj, Luo2026-ny, Yoshimochi2026-uh} and chiral magnon band splitting \cite{Sun2025-lz}. These phenomena arise from the characteristic symmetry of hematite, namely, its time-reversal-odd ($\mathcal{T}$-odd) magnetic structure. This symmetry of hematite has been discussed in the context of both \textit{g}-wave altermagnetism \cite{Galindez-Ruales2025-ja, Sun2025-lz, Smejkal2022-bn, Song2025-bl} and multipolar order parameters \cite{Verbeek2023-sx, Verbeek2024-rh}. Consequently, hematite serves as an excellent platform for exploring phenomena associated with the breaking of $\mathcal{T}$-symmetry.

Although the magnetic structures of hematite below and above the Morin transition temperature $T_\mathrm{M}$ are (nearly) collinear, the interplay with the crystal symmetry leads to a highly nontrivial order parameter. Specifically, high-rank magnetic multipoles—including octupoles (rank 3), hexadecapoles (rank 4), and triakontadipoles (rank 5) —have been proposed as the relevant degrees of freedom \cite{Verbeek2024-rh}. This implies that the underlying magnetic ordering is significantly more complex than a simple collinear picture suggests. Such hidden complexity is expected to be common in $\mathcal{T}$-odd antiferromagnets of sufficiently high point-group symmetry \cite{Bhowal2024-rt,Hariki2024-bq,Sasabe2025-dn}. Therefore, clarifying how such higher-order multipoles manifest in measurable physical quantities is a key to bridging the gap between symmetry-based classifications and experimentally observable phenomena.

An effective approach to accessing $\mathcal{T}$-odd multipole degrees of freedom is the observation of x-ray dichroism. There are different types of x-ray dichroism, including x-ray magnetic circular dichroism (XMCD), x-ray magnetochiral dichroism (XMChD), and x-ray nonreciprocal linear dichroism (XNLD). Each x-ray dichroism is sensitive to specific magnetic multipole orderings \cite{Di-Matteo2005-jg, Di-Matteo2012-tt,Lovesey1996-aa,Lovesey1996-bd}: XMCD primarily probes magnetic dipole \cite{Schutz1987-aj, Stohr1999-pm} and magnetic octupole \cite{Yamasaki2020-ow, Kimata2021-zu}, XMChD is sensitive to magnetic toroidal dipole \cite{Di-Matteo2005-jg,Ceolin2012-wx}, and XNLD reflects magnetic quadrupole \cite{Di-Matteo2005-jg, Kubota2004-oz}. In addition to this selectivity, measurements of x-ray dichroism offer two key advantages over visible and near-infrared optical approaches: element specificity and the capability to make direct comparisons with theoretical calculations.

However, the aforementioned x-ray dichroism measurements suffer from a limitation. When multipoles are ordered in a compensated antiferroic manner, local dichroic signals cancel each other out, preventing the observation of macroscopic dichroism.  To overcome such a limitation, we propose x-ray dichroism measurements of $\mathcal{T}$-odd antiferromagnets under an applied electric field (\textit{E}) which explicitly breaks space-inversion ($\mathcal{P}$) symmetry. By driving the system into a $\mathcal{P}$-odd and $\mathcal{T}$-odd state, it becomes possible not only to detect $\mathcal{T}$-odd properties through nonreciprocal responses but also to access ‘hidden’ antiferroic multipole orders. Recently, \textit{E}-induced magnetic toroidal dipoles in $\mathcal{T}$-odd antiferromagnets have been theoretically discussed \cite{Schmid2001-xr, Hayami2023-ok} and experimentally investigated via \textit{E}-induced nonreciprocal directional dichroism in the visible to near-infrared regions \cite{Hayashida2025-on, Kobayashi2026-ys}; however, a microscopic understanding of the specific multipole contributions remains elusive.

To demonstrate this proposal, we investigated x-ray dichroism near the Fe \textit{K}-edge of hematite under an applied \textit{E}. Our measurements revealed XNLD scales linearly with \textit{E} above $T_\mathrm{M}$. Furthermore, detailed polarization analysis and domain-dependent measurements, integrated with theoretical calculations, demonstrate that the observed \textit{E}-induced XNLD originates from the components of magnetic toroidal octupole and magnetic quadrupole.

This paper is organized as follows: In Sec.~\ref{sec:Symmetry}, we introduce the magnetic structure of hematite and discuss expected x-ray dichroism under \textit{E} in the perspective of symmetry. In Sec.~\ref{sec:Methods}, experimental and calculation methods are introduced. In Sec.~\ref{sec:Results}, we show experimental results. In Sec.~\ref{sec:Discussion}, we discuss the obtained results using tensor-based analysis and numerical simulations. Finally, in Sec.\ref{sec:Summary}, we summarize our work.

\section{\label{sec:level1} SYMMETRY ASPECT OF ELECTRIC-FIELD-INDUCED X-RAY NONRECIPROCAL DICHROISM IN HEMATITE}
\label{sec:Symmetry}

\subsection{\label{sec:level2}Magnetic structure of hematite}

\begin{figure}[t]
  \centering
  \includegraphics[width=0.9\columnwidth]{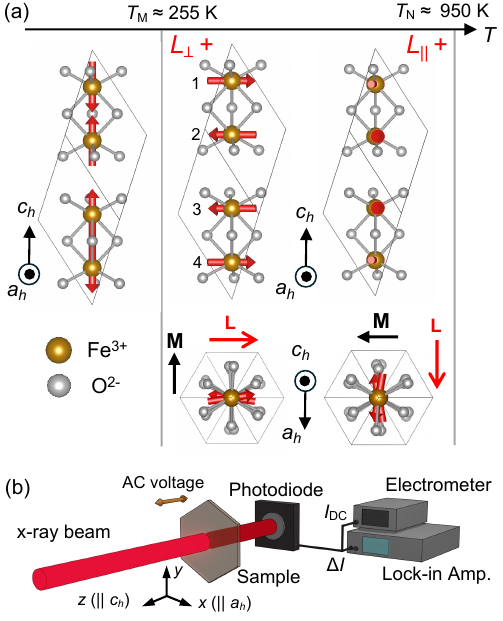}
  \caption{
  (a) Magnetic structures of hematite viewed along the $a_h$ axis (top panels) and the $c_h$ axis (bottom panels). Thick red arrows denote Fe spins, and black lines represent the rhombohedral primitive cell. The numbers next to the Fe$^{3+}$ ions denote the site numbers. Above $T_\mathrm{M}$, the Fe spins lie perpendicularly to the $c_h$ axis. The two states of $L_{\perp}+$ and $L_{\parallel}+$, with the Néel vector $\mathbf{L}$ aligned perpendicular and parallel to the $a_h$ axis, respectively, are depicted. $\mathbf{M}$ denotes net magnetization that appears due to spin canting.  Below $T_\mathrm{M}$, the spins are aligned along the $c_h$ axis. (b) Schematic illustration of the experimental setup.
  }
  \label{fig:structure}
\end{figure}

Hematite crystallizes in the corundum structure with the space group $R\bar{3}c$ \cite{Morish1994-nw, Pauling1925-vh}. Below $T_\mathrm{N} \approx 950~\mathrm{K}$, it exhibits a weakly canted AFM order, where spins lie in the basal $(001)_h$ plane ($h$ in the subscript denotes the hexagonal setting) [see right panel of Fig.~\ref{fig:structure}(a)] \cite{Morin1950-dl, Morish1994-nw}. The Néel vector is defined as $\mathbf{L} = \mathbf{M}_1 - \mathbf{M}_2 - \mathbf{M}_3 + \mathbf{M}_4$, where $\mathbf{M}_i$ (\textit{i} = 1 - 4) are Fe spins of site $i$ [for the site labels, see Fig.~\ref{fig:structure}(a)]. When $\mathbf{L}$ is perpendicular to the $a_h$ axis (termed as $L_{\perp}$ state), the magnetic point group is $2/m$ \cite{Harrison2010-fi, Oravova2013-al}. In the $L_{\perp}$ state, the Fe spins are slightly canted due to the DM interaction, resulting in a net magnetization along the $a_h$ axis. The two domain states related by the $\mathcal{T}$ operation, corresponding to a reversal of all spins, are termed as $L_{\perp}+$ and $L_{\perp}-$. $\mathbf{L}$ can freely rotate in the basal plane by applying an in-plane magnetic field due to the spin canting. When $\mathbf{L}$ is parallel to the $a_h$ axis, the magnetic point group is $2^{\prime}/m^{\prime}$ (termed as $L_{\parallel}$ state) \cite{Oravova2013-al}. As for the $L_{\perp}$ state, the two domain states related by the $\mathcal{T}$ operation are termed as $L_{\parallel}+$ and $L_{\parallel}-$. At $T_\mathrm{M} \approx 255~\mathrm{K}$, hematite undergoes the Morin transition, where the Fe spins are aligned parallel to the $c_h$ axis with a down-up-up-down configuration [see left panel of Fig.~\ref{fig:structure}(a)] \cite{Morin1950-dl}. The magnetic point group of this low-temperature ($T$) phase ($T \le T_{\mathrm{M}}$) is  $\bar{3}m$ \cite{Verbeek2024-rh, Oravova2013-al}, which also breaks $\mathcal{T}$ symmetry despite the absence of net magnetization. Although our main focus is the high-$T$ phases ($T_{\mathrm{M}} \le T \le T_{\mathrm{N}}$), we also discuss the low-$T$ phase for comparison.

\subsection{\label{sec:level2}X-ray nonreciprocal dichroism under an electric field}
\label{sec:sym_dichro}

Before discussing x-ray dichroism under \textit{E}, we briefly summarize previous x-ray measurements in the high-$T$ phases of hematite. Because of in-plane magnetic anisotropy, x-ray linear dichroism (XLD) is expected, and has been observed experimentally at the Fe \textit{L}-edge  \cite{Kuiper1993-wz, Suturin2021-bw}. We note that XLD is sensitive to the anisotropy of the square of magnetization and therefore cannot distinguish the domain states of $L+$ and $L-$. Since there is a net magnetization, XMCD is also expected, although it has been reported as a very small effect \cite{Suturin2021-bw}. However, very recently, XMCD originating from altermagnetic order has been reported \cite{Ishii2026-dx, Yamamoto2026-gy}. Magnetic properties of hematite have also been investigated by the resonant x-ray scattering, and the presence of local $\mathcal{P}$-odd/$\mathcal{T}$-even \cite{Kokubun2008-oe} and $\mathcal{P}$-odd/$\mathcal{T}$-odd multipoles has been suggested \cite{Lovesey2011-dj,Rodriguez-Fernandez2013-rd}.

We now turn to the \textit{E}-induced x-ray dichroism in a general framework, specifically focusing on the component that is linear in \textit{E}. In this case, $\mathcal{P}$-symmetry is explicitly broken by \textit{E}. Combined with the intrinsic breaking of $\mathcal{T}$-symmetry, the induced x-ray dichroism becomes $\mathcal{P}$-odd and $\mathcal{T}$-odd. Among different types of x-ray dichroism, XMChD and XNLD belong to this class \cite{Di-Matteo2012-tt}. Both XMChD and XNLD correspond to x-ray nonreciprocal directional dichroism (XNDD), in which absorption differs for counter-propagating beams. XMChD refers to the XNDD for unpolarized light, whereas XNLD describes XNDD for linearly polarized light. Note that although MChD is often used to denote dichroism in magnetized chiral materials, here we use this term more generally to describe NDD observed with unpolarized light. In XNLD, the linear polarization dependence reverses upon inversion of the propagation direction ($+k$ $\rightarrow$ $-k$).

To classify the symmetry conditions that host XNDD, it is convenient to introduce the linear magnetoelectric (ME) tensor, because the linear ME effect also occurs in $\mathcal{P}$-odd and $\mathcal{T}$-odd materials \cite{Spaldin2008-th,Spaldin2013-tu,Hayami2018-eu}. The linear ME effect is described as $P_i = \alpha_{ij}H_j$, where $P_i$ is polarization, $H_j$ is a magnetic field, and $\alpha_{ij}$ is the linear ME tensor that depends on the magnetic point group of the material \cite{Birss1966-vq}. Here, we consider the case where the x-ray beam propagates along the $z$ axis. In this geometry, XNDD can appear when the off-diagonal ME tensor elements $\alpha_{xy}$ and $\alpha_{yx}$ are nonzero. When the tensor is symmetric, i.e., $\alpha_{xy}= \alpha_{yx}$, XNLD can be nonzero, whereas when it is antisymmetric, i.e., $\alpha_{xy}= -\alpha_{yx}$, XMChD can be nonzero. Further details of the relationship between the ME tensor and XNDD are discussed in Appendix \ref{sec:MEtensor}.

Finally, we discuss \textit{E}-induced NDD in hematite from a symmetry perspective. To classify the possible responses, we consider the quadratic (second-order) ME effect, described as $P_i = \beta_{ijk}H_jE_k$. Here, $\beta_{ijk}$ is a third-rank $\mathcal{T}$-odd axial tensor that belongs to the same tensor class as the piezomagnetic tensor \cite{Birss1966-vq, Newnham2005-fg}. This can be regarded as an induced linear ME tensor, with $\alpha_{ij} = \beta_{ijk}E_k$. In the $L_{\perp}+$ state with point group $2/m$, when the x-ray beam propagates along the $z$ ($\parallel c_h$) axis and \textit{E} is applied along the same axis, the induced linear ME tensor can contain finite off-diagonal components $\alpha_{xy}$ and $\alpha_{yx}$ (with $x\parallel a_h$ and $y\parallel a_h\times c_h$), where $\alpha_{xy}\neq \alpha_{yx}$. This indicates that, in the configuration $\mathbf{E}\parallel \mathbf{k}\parallel c_h$, both XMChD and XNLD are allowed. On the other hand, in the $L_{\parallel}$ state with the magnetic point group is $2^{\prime}/m^{\prime}$, the ME tensor induced by $E_z$ has vanishing off-diagonal components ($\alpha_{xy}=\alpha_{yx}=0$), suggesting both XMChD and XNLD for $x$- or $y$- polarized beam are forbidden. However, XNLD is expected to be allowed, and in fact maximized, for linearly polarized x-ray beams oriented at $\pm 45^\circ$ with respect to the original $x$ axis when the coordinate system is rotated by $45^\circ$ to new $(x^\prime, y^\prime)$, where the tensor exhibits symmetric off-diagonal components  $\alpha_{x^\prime y^\prime}= \alpha_{y^\prime x^\prime}=(\alpha _{xx}-\alpha _{yy})/2$. In the low-$T$ phase with point group $\bar{3}m$, the induced ME tensor components under the same condition satisfy $\alpha_{xy} = –\alpha_{yx}$, suggesting that only XMChD is allowed. We emphasize that this ME tensor analysis determines which types of x-ray dichroism can appear but does not provide information on their magnitude.

\section{\label{sec:level1}METHODS}
\label{sec:Methods}

\subsection{\label{sec:level2}Experiments}

A natural single crystal of hematite, with the widest face parallel to the $(001)_h$ plane, was purchased from SurfaceNet GmbH. Magnetization measurements confirmed that the crystal exhibits a spontaneous magnetization at room temperature and undergoes the Morin transition at $T_\mathrm{M} \approx 253~\mathrm{K}$ [see Fig. \ref{fig:mag.prop.}(a) in Appendix~\ref{sec:mag.prop.}]. The sample was polished down to a thickness of $50\,\mu\mathrm{m}$. To apply an electric field along the thickness direction, indium-tin oxide (ITO) electrodes were sputtered onto both surfaces.
X-ray absorption measurements using photon energies near the Fe \textit{K}-edge were performed at the 19LXU beamline in SPring-8, Japan \cite{Yabashi2001-hf}. Figure ~\ref{fig:structure}(b) shows a schematic illustration of the experimental setup. Here, we define the coordinate system such that the $x$ axis is parallel to $a_h$, the $z$ axis to $c_h$, and the $y$ axis perpendicular to both. The linearly polarized x-ray beam propagated along the $z$ axis, and the DC component of transmitted intensity ($I_{\mathrm{DC}}$) was detected using a photodiode and an electrometer. For the measurements of \textit{E}-induced x-ray dichroism, we employed a lock-in technique. An AC voltage with a frequency of 999 Hz was applied to the sample along the $z$ axis using a function generator (WF1968, NF Corp.) and a voltage amplifier (A400, Pendulum Instr.). The AC component of the transmitted x-ray intensity ($\Delta I$), oscillating at the same frequency as the applied AC voltage, was measured using a lock-in amplifier (LI5645, NF Corp.). $\Delta I$ denotes the RMS (root-mean-square) value of the AC signal. For example, when an AC voltage with a peak amplitude of 200 V is applied, the observed $\Delta I$ corresponds to the change in the transmitted x-ray intensity by applying $200/\sqrt2 = 141.4\, \mathrm{V}$ ($E = 28.3 \, \mathrm{kV/cm}$). A magnetic field was applied by using a pair of neodymium magnets. Unless otherwise mentioned, the measurements were performed at room temperature in the high-$T$ phases.

\subsection{\label{sec:level2}Numerical simulations}
\label{sec:Methods_simulations}

For numerical simulations of the \textit{E}-induced XNDD in hematite, we employed the FDMNES \cite{Bunau2009-id} code, which, for this study, uses the density functional theory (DFT). To solve the electronic structure, which is mandatory step in our \textit{ab initio} approach, FDMNES can work in the following two modes before the calculation of the absorption cross section. The first mode uses the finite difference method (FDM), which has the advantage of being a full potential approach. The second mode uses the multiple scattering theory (MST) which may be less precise but is less CPU-intensive. Most of the results presented here were obtained using the MST. Another benefit of FDMNES is that it allows for the analysis of material characteristics and of the resulting spectroscopic signal in terms of $\mathcal{P}$-odd and $\mathcal{T}$-odd multipoles, as mentioned in the previous sections. Their components are expressed in a spherical tensor basis for all possible transition channels including electric dipole - electric dipole (E1-E1), electric dipole - electric quadrupole (E1-E2), and E2-E2. These serve as the basis for the analysis of the experimental data in the present paper, as shown in Sec.~\ref{sec:multipoleanalysis}. Further details of the simulations are given in Appendix \ref{sec:detailsofsim}.

\section{\label{sec:level1}RESULTS}
\label{sec:Results}

\subsection{\label{sec:level2}Electric-field dependence}

\begin{figure}[b]
  \centering
  \includegraphics[width=0.9\columnwidth]{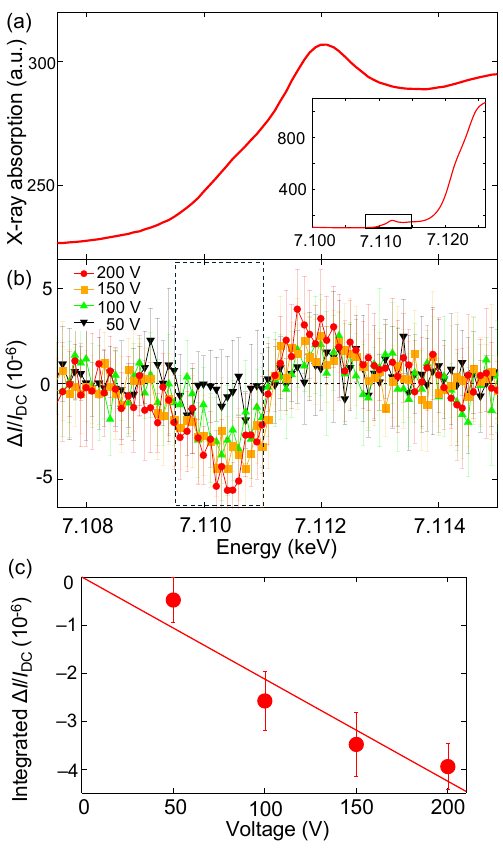}
  \caption{
   Spectra of x-ray absorption and electric-field-induced change in x-ray transmittance. (a) X-ray absorption spectrum around the Fe \textit{K} pre-edge. The inset shows the spectrum in the wide energy range including the main edge. (b)  Electric-field-modulated x-ray transmittance spectrum. The AC component of the transmitted light intensity ($\Delta I$), obtained under an applied AC voltage, is normalized by the DC component ($I_{\mathrm{DC}}$). The spectra obtained with the peak amplitude of 200 V, 150 V, 100 V, and 50 V are shown. (c) Voltage dependence of $\Delta I/I_{\mathrm{DC}}$ integrated in the energy range from 7.1095 keV to 7.1110 keV [see the dashed-line region in panel (b)]. The red line shows the result of least-square fitting.
  }
  \label{fig:XAS_Edep}
\end{figure}

Figure~\ref{fig:XAS_Edep}(a) shows the x-ray absorption spectrum around the Fe \textit{K}-edge.  The inset displays the spectrum over a wide energy range including both the pre-edge and main-edge regions, while the main panel presents a magnified view of the pre-edge peak. The obtained absorption spectrum agrees well with those in previous studies \cite{Kokubun2008-oe,Finkelstein1992-gf}. The pre-edge peak is attributed to the $1s$ $\rightarrow$ $3d$ E2 transition \cite{Kokubun2008-oe}, which provides the dominant contribution to XNDD \cite{Di-Matteo2012-tt, Lovesey1996-bd}. Therefore, we focus on this specific energy range.

We now present the results of the x-ray transmittance measurements under \textit{E}. During the measurements, magnetic fields of 0.1 T were applied along the $x$ axis to prepare a monodomain state of $L_{\perp}+$ [see Fig. \ref{fig:mag.prop.}(b) in Appendix~\ref{sec:mag.prop.}]. The incident x-ray beam was linearly polarized along the $x$ axis. An AC voltage with a peak amplitude of 200 V was applied, and the transmitted x-ray intensity modulated at the same frequency was detected. Figure~\ref{fig:XAS_Edep}(b) shows the resulting spectra of the AC component ($\Delta I$) normalized by the DC one ($I_{\mathrm{DC}}$), $\Delta I/I_{\mathrm{DC}}$. Although the obtained changes were small ($\Delta I/I_{\mathrm{DC}}$ is on the order of $10^{-6}$), the spectra exhibit a clear dispersive line shape. This shape is characterized by two peaks of opposite signs at 7.1105 keV and 7.1120 keV, with a zero crossing between them.

Figure~\ref{fig:XAS_Edep}(b) also shows the $\Delta I/I_{\mathrm{DC}}$ spectra measured under AC voltages with peak amplitudes of 150, 100, and 50 V. The $\Delta I/I_{\mathrm{DC}}$ spectrum systematically increases with increasing voltage. Figure~\ref{fig:XAS_Edep}(c) displays the voltage dependence of $\Delta I/I_{\mathrm{DC}}$ integrated in the energy range of 7.1095 keV - 7.1110 keV. The integrated signal is linear to the applied voltage. These results demonstrate that the high-$T$ phase of hematite exhibits the \textit{E}-induced XNDD expected from its symmetry. 

\begin{figure*}[t]
  \centering
  \includegraphics[width=0.75\textwidth]{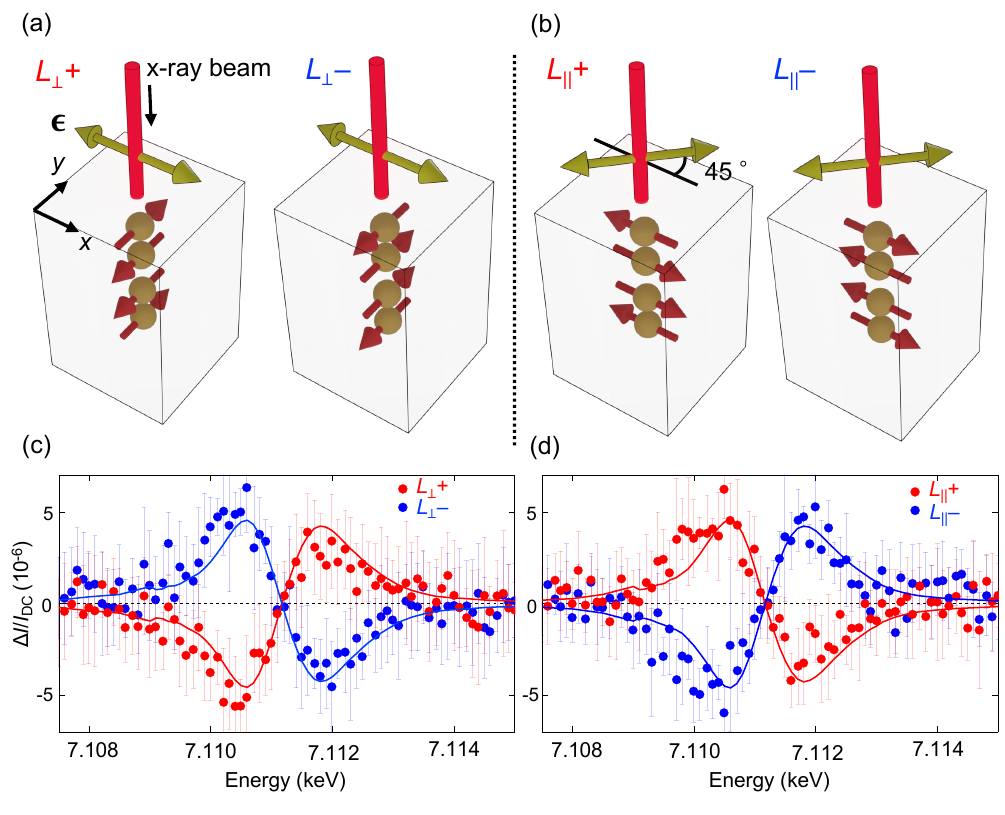}
  \caption{
   Domain state dependence of electric-field–modulated x-ray transmittance. (a,b) Schematic illustration of the experimental geometry for the $L_{\perp}$ (a) and $L_{\parallel}$ (b) states. The polarization of the incident x-ray beam ($\boldsymbol{\epsilon}$) is aligned along the $x$ axis for measurements of the $L_{\perp}\pm$ states and oriented at $45^\circ$ with respect to the $x$ axis for the measurements of the $L_{\parallel}\pm$ states. (c,d) Electric-field–modulated x-ray transmittance spectrum in the $L_{\perp}$ (c) and $L_{\parallel}$ (d) states. Measurements were performed in the $L_{\perp}+$ and $L_{\parallel}+$ (red), and $L_{\perp}-$ and $L_{\parallel}-$ (blue) domains. The spectra were obtained under an applied AC voltage with a peak amplitude of 200 V. Solid curves represent the results of numerical simulations (see Sec. VB).
  }
  \label{fig:XAS_domaindep}
\end{figure*}

\subsection{\label{sec:level2}Domain state dependence}

Next, we examined the domain state dependence of \textit{E}-induced XNDD. To prepare monodomain states of $L_{\perp}\pm$ (or $L_{\parallel}\pm$), magnetic fields of $\pm 0.1\,\mathrm{T}$ were applied along the $x$ (or $y$) axis during the measurements. The applied AC voltage was set at a peak amplitude of 200 V. For the measurements of the $L_{\perp}$ state, the incident x-ray was linearly polarized along the $x(\parallel a_h)$, i.e., with electric field of the x-ray $\boldsymbol{\epsilon} \parallel x$ [Fig.~\ref{fig:XAS_domaindep}(a)]. As shown in Fig.~\ref{fig:XAS_domaindep}(c), the spectrum reverses its sign upon switching between the $L_{\perp}+$ and $L_{\perp}-$ domains [compare red and blue dots in Fig.~\ref{fig:XAS_domaindep}(c)], indicating the $\mathcal{T}$-odd nature of the phenomenon. We also confirmed that the same spectrum was obtained after removing the magnetic field (see Fig.~\ref{fig:zerofield} in Appendix \ref{sec:zerofield}), demonstrating that the AFM order, rather than the small net magnetic moment, provides the dominant contribution. Furthermore, in the $L_{\parallel}+$ state, \textit{E}-induced XNDD with a magnitude comparable to that in the $L_{\perp}$ state was observed for linearly polarized x-ray oriented at $45^\circ$ with respect to the $x$ axis [Figs.~\ref{fig:XAS_domaindep}(b) and \ref{fig:XAS_domaindep}(d)], which is consistent with the symmetry-based expectation discussed in Sec.~\ref{sec:sym_dichro}.

Although the propagation direction of the x-ray beam was not reversed in the present experiment, the nonreciprocal nature of the observed response is confirmed by its linear dependence on the applied \textit{E} and the sign reversal between the domain states related by $\mathcal{T}$ operation. In these measurements, an AC electric field was applied, and lock-in detection synchronized to the AC voltage extracts the signal component proportional to \textit{E}. Consequently, the observed $\Delta I$ reverses sign upon effective inversion of \textit{E} (i.e., across the positive and negative half-cycles), providing direct evidence of the $\mathcal{P}$-odd nature. This equivalence arises from the symmetry of the system: reversing the beam propagation direction ($+k$~$\rightarrow$~$-k$) under fixed domain and \textit{E} is equivalent to reversing the sign of either the domain state ($\mathcal{T}$-odd) or the applied \textit{E} ($\mathcal{P}$-odd), both of which reverse the sign of the nonreciprocal effect.

\subsection{\label{sec:level2}Polarization angle dependence}

\begin{figure*}[t]
  \centering
  \includegraphics[width=0.85\textwidth]{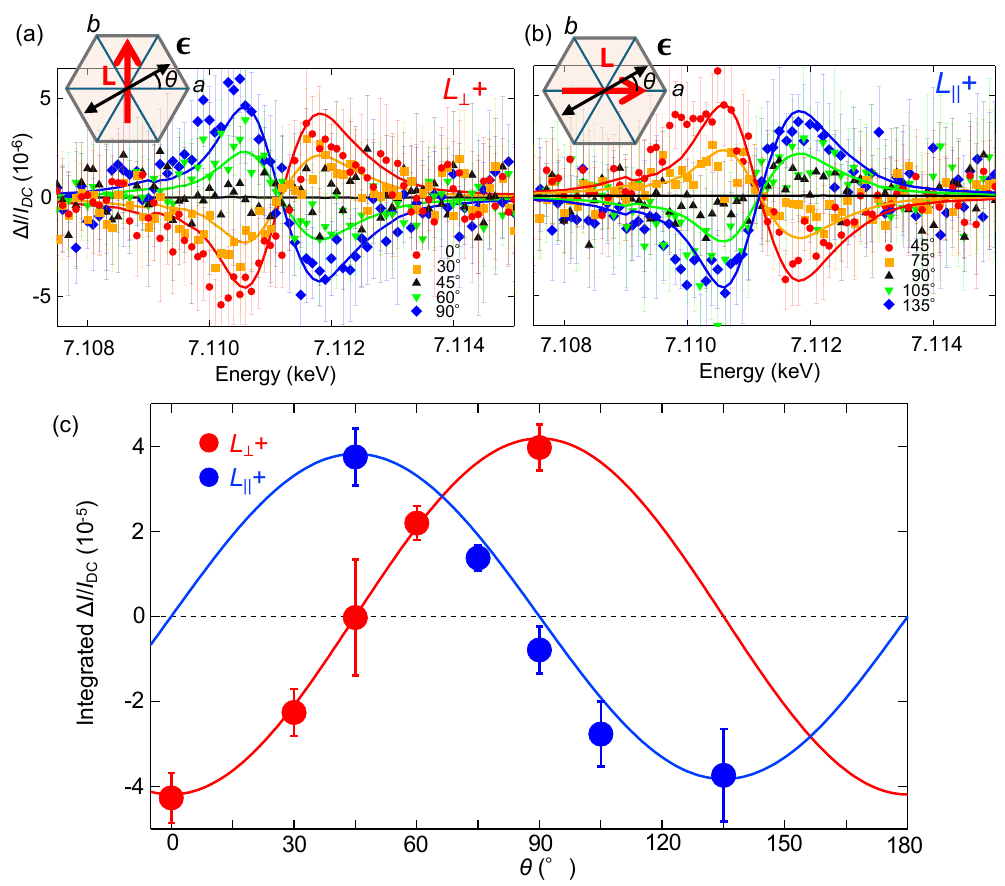}
  \caption{
  Polarization angle dependence of electric-field–modulated x-ray transmittance. (a,b) Electric-field–modulated x-ray transmittance spectra measured for different polarization angles ($\theta$) in the $L_{\perp}+$ domain (a) and $L_{\parallel}+$ domain (b). The spectra were obtained under an applied AC voltage with a peak amplitude of 200 V. Solid curves represent the results of numerical simulations (see Sec.~\ref{sec:simulations}). (c) Polarization angle dependence of $\Delta I/I_{\mathrm{DC}}$ integrated in the energy range from 7.1095 keV to 7.1110 keV for the $L_{\perp}+$ domain (red) and $L_{\parallel}+$ domain (blue). Solid curves represent the results of least-squares fits using a cosine function ($\propto \cos 2\theta$) for the $L_{\perp}+$ domain and a sine function ($\propto \sin 2\theta$) for the $L_{\parallel}+$ domain.
  }
  \label{fig:XAS_poldep}
\end{figure*}

Next, to identify the \textit{E}-induced XNDD that originates from XMChD, XNLD, or both, we investigated the dependence of $\Delta I/I_{\mathrm{DC}}$ on the in-plane rotation of $\boldsymbol{\epsilon}$. Figure~\ref{fig:XAS_poldep}(a) shows the $\Delta I/I_{\mathrm{DC}}$ spectra measured at the polarization angle of $\theta = 0^\circ,\, 30^\circ,\, 45^\circ,\, 60^\circ$, and $90^\circ$ in the $L_{\perp}+$ domain. Here, the origin of $\theta$ is set at $\boldsymbol{\epsilon}\parallel x$.  The measurements were performed under an AC voltage with a peak amplitude of 200 V. The spectra exhibit a complete sign reversal between linear polarizations parallel ($\theta = 0^\circ$) and perpendicular ($\theta = 90^\circ$) to the $x$ axis [compare red circles and blue diamonds in Fig.~\ref{fig:XAS_poldep}(a)]. When the polarization angle is intermediate ($\theta = 45^\circ$), the \textit{E}-induced NDD is strongly suppressed [black triangles in Fig.~\ref{fig:XAS_poldep}(a)]. This polarization angle dependence indicates that the \textit{E}-induced XNDD is dominated by XNLD, with any contribution from XMChD being negligible. This is evidenced by the complete sign reversal between $\theta = 0^\circ$ and $90^\circ$, near disappearance at $\theta = 45^\circ$, and the absence of a significant constant component across angles.  We performed the same measurements for the $L_{\parallel}+$ domain, prepared by applying a magnetic field of 0.1 T along the $y$ axis ($\perp a_h$). Figure~\ref{fig:XAS_poldep}(b) shows the corresponding spectra. \textit{E}-induced XNDD is also observed in the $L_{\parallel}+$ domain. However, the signals are maximized at $\theta = 45^\circ$ and $\theta = 135^\circ$ but with opposite phases. These results show that when the $\mathbf{L}$ vector is rotated by $\theta = 90^\circ$, the phase of the \textit{E}-induced XNLD is shifted by $\theta = 135^\circ$ (or equivalently $\theta = -45^\circ$). This phase shift is evident in the $\theta$ dependence of the integrated $\Delta I/I_{\mathrm{DC}}$ shown in Fig.~\ref{fig:XAS_poldep}(c). The origin of this phase shift between the $L_{\perp}$ and $L_{\parallel}+$ states will be discussed below.

\section{\label{sec:level1}DISCUSSION}
\label{sec:Discussion}

Here, we discuss the origin of the observed \textit{E}-induced XNDD spectra. We first present a tensor-based analysis that captures the essential features of the results, especially focusing on the polarization angle dependence. Then we discuss a microscopic interpretation based on \textit{ab initio} DFT calculations.

\subsection{\label{sec:level2}Tensor-based analysis}
\label{Sec:tensorAna}
The general x-ray absorption process involving E1 and E2 transitions is described as
\begin{align}
\sigma(\mathcal{E}) \propto
\sum_n
\Big|
\left\langle \Psi_0 \middle|
\bm{\epsilon}\cdot\mathbf{r}
\left(1+\frac{i}{2}\mathbf{k}\cdot\mathbf{r}\right)
\middle| \Psi_n
\right\rangle
\Big|^2
\nonumber \\
\delta\!\left(\mathcal{E}-(\mathcal{E}_n-\mathcal{E}_0)\right)
\label{eq:absorption}
\end{align}
where $\Psi_0$ and $\Psi_n$ are the ground and excited states with energies $\mathcal{E}_0$ and $\mathcal{E}_n$, respectively. Here, $\bm{\epsilon}$ is the electric field of the x-ray, $\mathbf{k}$ is the propagation direction, $\mathbf{r}$ is the position operator, and $\mathcal{E}$ denotes photon energy. Because the E1-E1 and E2-E2 transition processes in Eq.~(\ref{eq:absorption}) are the $\mathcal{P}$-even [45], XNDD, defined as $\sigma_{dic}=\sigma_{k+}-\ \sigma_{k-}$, originates purely from the E1-E2 transition. It can be written as
\begin{align}
\sigma_{\mathrm{dic}}(\mathcal{E}) \propto
&i\,\epsilon_\alpha k_\beta \epsilon_\gamma
\sum_n
\Big[
\langle \Psi_0 | r_\alpha r_\beta | \Psi_n \rangle
\langle \Psi_n | r_\gamma | \Psi_0 \rangle
\nonumber \\
&-
\langle \Psi_0 | r_\alpha | \Psi_n \rangle
\langle \Psi_n | r_\beta r_\gamma | \Psi_0 \rangle
\Big]
\delta\!\left(\mathcal{E}-(\mathcal{E}_n-\mathcal{E}_0)\right)
\nonumber \\
&=
i\,\epsilon_\alpha k_\beta \epsilon_\gamma
T_{\alpha\beta\gamma}\,
\label{eq:dichroism}
\end{align}
where repeated Cartesian indices imply summation. The matter tensor $T_{\alpha\beta\gamma}$ in Eq.~(\ref{eq:dichroism}) is defined as
\begin{equation}
T_{\alpha\beta\gamma}
=
\sum_n
\Big[
\langle r_\alpha r_\beta \rangle_{0n}
\langle r_\gamma \rangle_{n0}
-
\langle r_\alpha \rangle_{0n}
\langle r_\beta r_\gamma \rangle_{n0}
\Big]\delta _n
\label{eq:mattertensor}
\end{equation}
and depends solely on the intrinsic properties of the material, subject to its magnetic point symmetry constraints. Here we define $\delta_n = \delta\!\left(\mathcal{E}-(\mathcal{E}_n-\mathcal{E}_0)\right)$. In general, $T_{\alpha\beta\gamma}$ is a complex tensor, whose real and imaginary parts correspond to $\mathcal{T}$-even and $\mathcal{T}$-odd properties, respectively. Since the present study concerns $\mathcal{T}$-odd responses, we focus on the imaginary components of $T_{\alpha\beta\gamma}$.

In our geometry (propagation direction $k_z$), XNDD takes the form:
\begin{align}
\sigma_{\mathrm{dic}}(\mathcal{E})
&\propto
i\Big[
\epsilon_x k_z \epsilon_x T_{xzx}
+\epsilon_y k_z \epsilon_y T_{yzy}
\nonumber \\
&\qquad
+\epsilon_x k_z \epsilon_y (T_{xzy}+T_{yzx})
\Big]
\nonumber \\
&=
i k_z
\Big[
\frac{1}{2}(\epsilon_x\epsilon_x+\epsilon_y\epsilon_y)(T_{xzx}+T_{yzy})
\nonumber \\
&\qquad
+\frac{1}{2}(\epsilon_x\epsilon_x-\epsilon_y\epsilon_y)(T_{xzx}-T_{yzy})
\nonumber \\
&\qquad
+\epsilon_x\epsilon_y(T_{xzy}+T_{yzx})
\Big].
\label{eq:dichroism_mattertensor}
\end{align}
Introducing the polarization angle $\theta$, the electric field components are described as $\epsilon_x=\epsilon\cos{\theta}$ and $\epsilon_y=\epsilon\sin{\theta}$, which leads to $\epsilon_x\epsilon_x+\epsilon_y\epsilon_y=\ \epsilon^2\cos^2{\theta}+\epsilon^2\sin^2{\theta}=\epsilon^2$, $\epsilon_x\epsilon_x-\epsilon_y\epsilon_y=\epsilon^2\cos^2{\theta}-\epsilon^2\sin^2{\theta}=\epsilon^2\cos{2\theta}$, and $\epsilon_x\epsilon_y=\frac{1}{2}\sin{2\theta}$. Accordingly, the term proportional to $T_{xzx}+T_{yzy}$ is independent of $\theta$ and contributes to \mbox{XMChD}, whereas $T_{xzx}-T_{yzy}$ and $T_{xzy}+T_{yzx}$ give rise to XNLD with the $\cos{2\theta}$ and $\sin{2\theta}$ dependence, respectively.

We now consider the symmetry-allowed components of $T_{\alpha\beta\gamma}$ in hematite under $E_z$. In the $L_{\perp}+$ state, characterized by the magnetic point group $2/m$, the application of $E_z$ reduces the symmetry to $m$, where the mirror plane is perpendicular to the $x$ axis (hereafter denoted as $m_x$). Under this symmetry, only $T_{\alpha\beta\gamma}$ containing even number of $x$ indices are allowed. Consequently, XMChD arising from $T_{xzx}+T_{yzy}$ and XNLD arising from $T_{xzx}-T_{yzy}$ are allowed by symmetry. Experimentally, however, only the XNLD component is observed, exhibiting a clear $\cos{2\theta}$ dependence [see the red curve in Fig.~\ref{fig:XAS_poldep}(c)].

In contrast, in the $L_{\parallel}$ state with the magnetic point group $2^{\prime}/m^{\prime}$, the symmetry under $E_z$ becomes $m_x^\prime$. In this case, only $T_{\alpha\beta\gamma}$ containing an odd number of $x$ indices are allowed. Consequently, XNLD arising from $T_{xzy}+T_{yzx}$ with a $\sin{2\theta}$ dependence is allowed. This is in excellent agreement with the experimentally observed polarization angle dependence in the $L_{\parallel}$ state [see the blue curve in Fig.~\ref{fig:XAS_poldep}(c)].

\subsection{\label{sec:level2}Microscopic analysis based on \textit{ab-initio} calculations}
\label{sec:simulations}

As mentioned in Sec.~\ref{sec:Methods_simulations}, we employed the FDMNES code to simulate and analyze the data. As a first step, x-ray absorption spectrum around the Fe \textit{K}-edge without applying an electric field was simulated and obtained the spectrum of the transmitted x-ray intensity $I(0)$. The calculated absorption spectrum shows good agreement with the experimentally observed one [compare Fig.~\ref{fig:sim_broad}(a) in Appendix~\ref{sec:detailsofsim} and the inset of Fig. ~\ref{fig:XAS_Edep}(a)].

To incorporate the effect of an applied electric field, we employed a simple model, assuming that the primary effect of the electric field could be a relative displacement of cations and anions along the $c_h$ axis. In practice, all Fe$^{3+}$ ions were uniformly shifted in the direction along $E$ ($\parallel c_h$), while all O$^{2-}$ ions were kept fixed. Accordingly, we calculated the normalized change in transmitted x-ray intensity, $\Delta I(\delta z)/I(0)$, by introducing the Fe displacement $\delta z$. The calculated spectra of $\Delta I(\delta z)/I(0)$ in the $L_{\perp}$ state are shown in Fig.~\ref{fig:XAS_domaindep}(c), as red and blue solid curves for the $L_{\perp}+$ and $L_{\perp}-$ domains, respectively. The simulated $\Delta I(\delta z)/I(0)$ spectra, which correspond to the experimental $\Delta I/I_{\mathrm{DC}}$ spectra, successfully reproduce both the dispersive line shape and the sign reversal between the two domains observed experimentally. The simulated spectra are mostly proportional to the Fe displacement $\delta z$. We found that the best agreement in the amplitude of $\Delta I(\delta z)/I(0)$ with the experimental $\Delta I/I_{\mathrm{DC}}$ (at 141.4 V, $E$ = 28.3 kV/cm) was obtained for a Fe displacement of $\delta z = 5 \times 10^{-15}\,\mathrm{m}$. We emphasize first that, apart from the Fe displacement, no additional adjustable parameters were introduced in the calculations. We also found it remarkable that XNDD is able to quantitatively determine the effect of the electric field on the material structure. Figure~\ref{fig:sim_broad}(b) in Appendix~\ref{sec:detailsofsim} shows the simulated spectra over a wide energy range, confirming that the \textit{E}-induced XNDD is most pronounced in the pre-edge region dominated by the $1s\rightarrow3d$ E2 transition

We further applied the same simulation framework to the polarization angle dependence in the $L_{\perp}+$ and $L_{\parallel}+$ domains. As indicated by the solid curves in Figs.~\ref{fig:XAS_poldep}(a) and \ref{fig:XAS_poldep}(b), the XNLD character, including the $-45^\circ$ phase shift between the two domains, is well reproduced by the calculations. The XMChD contribution, which is independent of the angle $\theta$, is negligibly small in the simulations, in good agreement with the experimental observations. 

We also calculated the $\Delta I(\delta z)/I(0)$ spectrum in the low-$T$ phase below $T_\mathrm{M}$, where \textit{E}-induced XMChD is symmetrically allowed. The calculation shows that the magnitude of $\Delta I(\delta z)/I(0)$ in the low-$T$ phase is about one order smaller than that in the high-$T$ phases [see Fig.~\ref{fig:lowT}(a) in Appendix~\ref{sec:lowT}]. Consistently, no meaningful $\Delta I/I_{\mathrm{DC}}$ signal was detected in the low-$T$ phase [Fig.~\ref{fig:lowT}(b) in Appendix~\ref{sec:lowT}], indicating that the signal was below the detection limit of the present experimental setup. These results strongly support that the observed \textit{E}-induced XNDD in hematite originates primarily from the XNLD mechanism associated with the high-$T$ phases.

\subsection{\label{sec:level2}Spherical-multipole analysis}
\label{sec:multipoleanalysis}

To obtain further insight into the physical meaning of the observed response, we decomposed the calculated \textit{E}-induced XNDD into spherical-multipole contributions. The matter tensor given in Eq.~(\ref{eq:dichroism_mattertensor}) and defined in Eq.~(\ref{eq:mattertensor}) can be rewritten in spherical components as
\begin{equation}
T_{xzx}+T_{yzy}
=
-\sqrt{\frac{2}{5}}
\left(
\sqrt{2}\, I_{3,0}
+
\sqrt{3}\, I_{1,0}
\right)
\label{eq:Mattertensor_1}
\end{equation}
\begin{equation}
T_{xzx}-T_{yzy}
=
\sqrt{\frac{1}{3}}
\left[
\sqrt{2}\,\left(I_{3,2}+I_{3,-2}\right)
-
\left(I_{2,2}-I_{2,-2}\right)
\right]
\label{eq:Mattertensor_2}
\end{equation}
\begin{equation}
T_{xzy}+T_{yzx}
=
\sqrt{\frac{1}{3}}
\left[
\sqrt{2}\,\left(I_{3,2}-I_{3,-2}\right)
-
\left(I_{2,2}+I_{2,-2}\right)
\right]
\label{eq:Mattertensor_3}
\end{equation}
where $I_{l,m}$, defined in Appendix \ref{sec:SphericalMultipole}, denotes $\mathcal{P}$-odd and $\mathcal{T}$-odd multipoles of rank $l$ with azimuthal quantum number $m$. The rank 1, 2, and 3 $\mathcal{P}$-odd and $\mathcal{T}$-odd multipoles are referred to as magnetic toroidal dipole, magnetic quadrupole, and magnetic toroidal octupole, respectively \cite{Di-Matteo2005-jg,Ramakrishnan2023-bu}. The correspondence between the spherical multipoles and the conventional Cartesian multipoles is given in Appendix \ref{sec:orbitalcurrent}. Equations~(\ref{eq:Mattertensor_1}) and (\ref{eq:Mattertensor_2}) describe the $L_{\perp}$ state whereas Eq.~(\ref{eq:Mattertensor_3}) is only active for the $L_{\parallel}$ state. 

\begin{figure}[t]
  \centering
  \includegraphics[width=0.9\columnwidth]{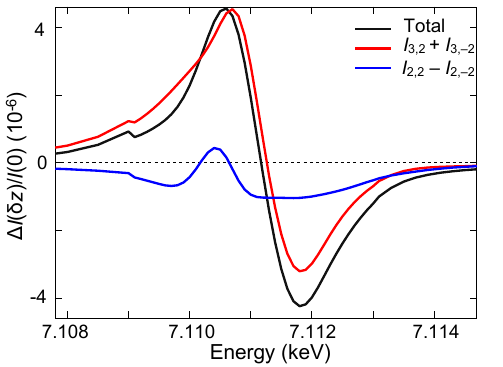}
  \caption{
  (a) Multipole decomposition of the simulated electric-field-induced XNLD spectrum. The total spectrum corresponds to the simulation for the $L_{\perp}$ domain shown in Fig.~\ref{fig:XAS_Edep}(b). The spectrum is decomposed into contributions from the magnetic toroidal octupole $I_{3,2}+I_{3,-2}$ and the magnetic quadrupole $I_{2,2}-I_{2,-2}$. The details of each multipole are discussed in the main text.
  }
  \label{fig:Multipole_decomp}
\end{figure}

Figure~\ref{fig:Multipole_decomp} shows the multipole decomposition of the simulated \textit{E}-induced XNDD spectrum. In the $L_{\perp}$ state, $\Delta I/I(0)$ is dominated by the contribution from the magnetic toroidal octupole $I_{3,2}+I_{3,-2}$ with a smaller but finite contribution from the magnetic quadrupole $I_{2,2}-I_{2,-2}$. In contrast, the toroidal dipole $I_{1,0}$ and the other toroidal octupole $I_{3,0}$ contribute negligibly. In the $L_{\parallel}$ state, the only allowed contributions are from the magnetic toroidal octupole $I_{3,2}-I_{3,-2}$ and a smaller $I_{2,2}+I_{2,-2}$ magnetic quadrupole, both rotated by $45^\circ$ in azimuth compared to the $L_{\perp}$ state, as experimentally found.

What does multipole analysis imply? Although the matter tensor $T_{\alpha\beta\gamma}$ must satisfy the symmetry constraints of the material, it is defined by the matrix elements connecting the ground and excited states. Thus, it primarily characterizes the response channel of the x-ray absorption process, rather than being identified directly with static multipoles of the entire system.

Nevertheless, for the Fe \textit{K}-edge considered here, the relevant transitions involve an isotropic $1s$ core state and E2 ($1s\rightarrow3d$) and E1 ($1s\rightarrow4p$) channels. Consequently, the symmetry information encoded in the matter tensor $T_{\alpha\beta\gamma}$ is largely governed by the $3d$ states, which also define the symmetry of the $\mathrm{Fe_2O_3}$ system under an applied electric field. From this perspective, the tensor/multipole analysis goes beyond mere classification of x-ray response channels and strongly suggests that the magnetic toroidal octupole and magnetic quadrupole constitute the dominant multipolar components characterizing the \textit{E}-induced state of hematite. Surprisingly, these \textit{E}-induced multipoles originate from extremely small displacements of Fe$^{3+}$ ions. Our precise measurements, together with detailed numerical analysis, reveal this subtle yet important property of the system. We also note that, since the $K$-edge transition from the spinless $1s$ initial state, the $\mathcal{T}$-odd properties appearing in Eq.~(\ref{eq:dichroism_mattertensor}) are introduced purely through the orbital degrees of freedom. Accordingly, the magnetic multipoles in Eq.~(\ref{eq:dichroism_mattertensor}) can be physically interpreted in terms of orbital currents (see Appendix \ref{sec:orbitalcurrent}).

Finally, we comment on the possible multipole ordering in the absence of an electric field. Because $\mathcal{P}$-symmetry is preserved, the $\mathcal{P}$-odd and $\mathcal{T}$-odd multipoles vanish upon averaging over the unit cell. However, the local site symmetry of individual Fe atoms is 3, which is $\mathcal{P}$-odd and $\mathcal{T}$-odd, allowing for the $\mathcal{P}$-odd and $\mathcal{T}$-odd multipoles, such as the magnetic toroidal octupole. The Fe sites form inversion-related pairs [sites 1 and 2, or 3 and 4 in Fig.~\ref{fig:structure}(a)], in which the $\mathcal{P}$-odd and $\mathcal{T}$-odd multipoles appear with opposite signs, leading to a compensated arrangement at zero electric field. Applying an electric field breaks this compensation and gives rise to a finite $\mathcal{P}$-odd and $\mathcal{T}$-odd multipolar response. From this viewpoint, the present study of \textit{E}-induced XNDD suggests that the high-$T$ phases of hematite inherently supports an antiferroic arrangement of magnetic toroidal octupoles and magnetic quadrupoles, which becomes macroscopically observable when the inversion-related compensation is lifted by an external electric field. 

\section{\label{sec:level1}SUMMARY}
\label{sec:Summary}

In conclusion, we have demonstrated that the AFM state of hematite above $T_\mathrm{M}$ exhibits \textit{E}-induced XNDD, primarily \textit{E}-induced XNLD. The observed spectra, including their polarization angle dependence and $\mathcal{T}$ domain sign reversal, are in excellent agreement with DFT-based simulations. Furthermore, our multipole analysis reveals that an electric field induces a magnetic toroidal octupole, suggesting an antiferroic ordering of these octupoles in the absence of an electric field. These results highlight that probing x-ray absorption as a response to an applied electric field provides a powerful means to access hidden symmetry information, surpassing the capabilities of conventional zero-field measurements.

Finally, the explicit identification of the magnetic toroidal octupole paves a pathway to exploring its associated physical consequences, including nonreciprocal transport phenomena arising from asymmetric electronic band structures \cite{Watanabe2020-lo,Yatsushiro2022-sf, Hayami2024-qk}.

\begin{acknowledgments}
The images of crystal structures were drawn using the software VESTA \cite{Momma2011-zf}. This work was supported by JSPS KAKENHI Grants Nos. JP24K22855, JP25H00392 and JP25H01247 and Murata Science and Education Foundation. X-ray experiments were performed at the BL19LXU beamline in SPring-8, with the approval from RIKEN (Proposals No. 20240005 and No.20250039).
\end{acknowledgments}

\appendix

\section{Linear magnetoelectric tensor and nonreciprocal directional dichroism}
\label{sec:MEtensor}

Here, we discuss the relationship between the linear ME tensor and nonreciprocal directional dichroism (NDD). We consider NDD of light in a general sense and do not restrict the discussion to the x-ray regime.

The relationship between the ME tensor and NDD can be understood by extending the ME effect to optical frequencies, often referred to as the optical ME effect \cite{Szaller2013-cu,Tokura2018-zz,Kimura2020-yu,Kimura2023-nv}. For simplicity, we consider light propagating along the $z$ direction, and thus in the following only the in-plane ($xy$) components of the ME tensor are considered. When the ME tensor in the $xy$ plane has finite off-diagonal components, oscillating polarization $\mathbf{P}^\omega$, which is normally induced by an oscillating electric field $\mathbf{E}^\omega$, acquires an additional contribution $\Delta \mathbf{P}^\omega$  induced by an oscillating magnetic field $\Delta \mathbf{H}^\omega$ of light. This additional polarization is described as
\begin{align}
\Delta \mathbf{P}^\omega
&=
\begin{pmatrix}
0 & \alpha_{xy}^{\omega} \\
\alpha_{yx}^{\omega} & 0
\end{pmatrix}
\mathbf{H}^\omega
\propto
\begin{pmatrix}
0 & \alpha_{xy}^{\omega} \\
\alpha_{yx}^{\omega} & 0
\end{pmatrix}
(\mathbf{k}\times\mathbf{E}^\omega)
\nonumber \\
&=
\pm
\begin{pmatrix}
\alpha_{xy}^{\omega} & 0 \\
0 & -\alpha_{yx}^{\omega}
\end{pmatrix}
\mathbf{E}^\omega .
\label{eq:OME}
\end{align}
Here, $\alpha_{xy}^\omega$ and $\alpha_{yx}^\omega$ are components of optical ME tensor. $\mathbf{k}$ is the unit vector indicating the propagation of light, and $k_z = \pm 1$ in the present case. As shown here, the sign of $\Delta \mathbf{P}^\omega$ depends on the propagation direction of light, giving rise to nonreciprocity. For simplicity and to highlight the essential origin of nonreciprocity, we set the diagonal components to zero here, as they alone do not produce a directional dependence in absorption upon reversing the light propagation direction.

When the off-diagonal components of the ME tensor are symmetric, i.e., $\alpha_{xy}^\omega=\alpha_{yx}^\omega$, the sign of $\Delta \mathbf{P}^\omega$ is reversed between $x$- and $y$-polarized light, leading to nonreciprocal linear dichroism (NLD). In contrast, when the off-diagonal components are antisymmetric, i.e., 
$\alpha_{xy}^\omega={-\alpha}_{yx}^\omega$, the sign of $\Delta \mathbf{P}^\omega$ remains the same for 
$x$- and $y$-polarized light, resulting in magnetochiral dichroism (MChD).
From this discussion, it is clear that the ME tensor determines which type of NDD can occur. On the other hand, as discussed in the main text, XNDD arises from electric quadrupole transitions, and the above picture based on polarization induced by an oscillating magnetic field does not necessarily reflect the actual mechanism in the x-ray regime. Nevertheless, this framework is useful for classifying and understanding the nonreciprocal optical phenomena from a symmetry perspective.

\begin{figure}[t]
  \centering
  \includegraphics[width=0.9\columnwidth]{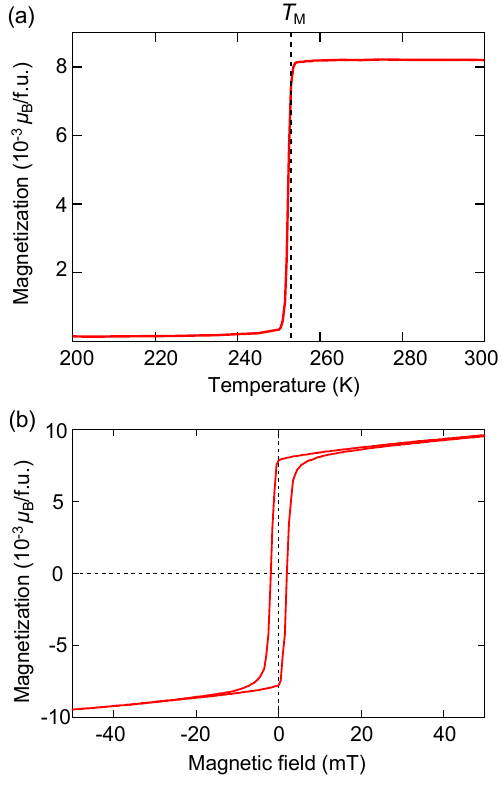}
  \caption{
  Magnetic properties of hematite. (a) Temperature dependence of magnetization along the $a_h$ axis. A magnetic field of 10 mT was applied during the measurement. (b) Magnetization curve at 300 K for a magnetic field applied along the $a_h$ axis.
  }
  \label{fig:mag.prop.}
\end{figure}

\section{Magnetic properties of hematite}
\label{sec:mag.prop.}

The magnetization of the natural hematite crystal was measured using a commercial magnetometer (PPMS, Quantum Design). Figure \ref{fig:mag.prop.}(a) shows the temperature dependence of the magnetization along the $a_h$ axis under a magnetic field of 10 mT. The sudden decrease in the magnetization at approximately 253 K corresponds to the Morin transition. Figure \ref{fig:mag.prop.}(b) shows the magnetization curve at 300 K for a magnetic field applied along the $a_h$ axis. A hysteresis loop with a coercive field of about $\pm$5 mT is observed.

\section{\textit{E}-induced XNDD measurements in zero magnetic field}
\label{sec:zerofield}
\begin{figure}[tb]
  \centering
  \includegraphics[width=0.9\columnwidth]{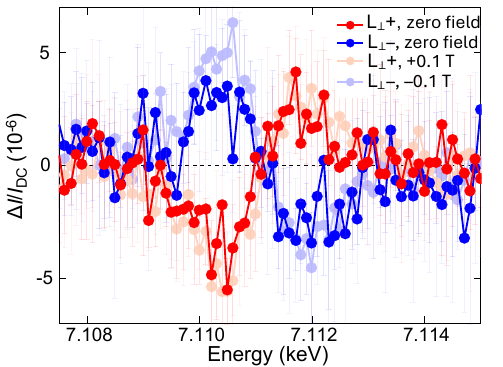}
  \caption{
   Electric-field-modulated x-ray transmittance spectrum in zero magnetic field. Monodomain states of $L_{\perp}+$ (red) and $L_{\perp}-$ (blue) were prepared by applying magnetic fields of +0.1 T and $-0.1\,\mathrm{T}$, respectively, and the magnetic field was removed during the measurements. The spectra were obtained under an applied AC voltage with a peak amplitude of 200 V. For reference, the spectra measured in magnetic fields [identical to those shown in Fig. \ref{fig:XAS_domaindep}(c)] are displayed in pale colors.
  }
  \label{fig:zerofield}
\end{figure}

Electric-field-modulated x-ray transmittance spectra were measured in zero magnetic field, under the same experimental conditions as those used for Fig. \ref{fig:XAS_domaindep}(c). To prepare monodomain states of $L_{\perp}+$ and $L_{\perp}-$, magnetic fields of +0.1 T and $-0.1\,\mathrm{T}$ were applied along the $a_h$ axis, respectively, and subsequently removed during the spectral measurements. An AC voltage with a peak amplitude of 200 V was applied, and the x-ray beam was linearly polarized along the $x$ axis. Figure \ref{fig:zerofield} shows the resulting spectra. No significant difference is observed between the spectra measured in zero magnetic field and those obtained under a finite magnetic field.

\section{Details of numerical simulations by FDMNES code}
\label{sec:detailsofsim}

Most of the calculations were performed using the Multiple scattering theory. We employed the Generalized Gradient Approximation (GGA) PBE96 \cite{Perdew1996-rz} functional to calculate the exchange-correlation potential, incorporating the correction by Hedin and Von Bart \cite{Barth1972-vr} to consider its dependence on the photoelectron energy. The hematite $R\bar{3}c$ structure was adopted from ref.~\cite{Cros1976-wb}, with hexagonal lattice parameters of $a_h = 5.035\,\text{\AA}$ and $c_h = 13.73\,\text{\AA}$. The atomic positions of Fe and O are (0, 0, 0.14783) and (0.30618, 0, 0.25), respectively. In the resulting unit cell, all Fe atoms were displaced along the $c_h$ axis by $\pm \delta z$, depending on the direction of the electric field. Two separate calculations were then performed for each field direction, and the results were compared with the experimental data by taking their differences. The magnetic configurations of $L_{\perp}$, $L_{\parallel}$, and the low-$T$ phase are imposed at the beginning of the calculation using the Hund rule. There is a very small canting, a fraction of degree, of the magnetic moments in the $L_{\perp}$ and $L_{\parallel}$ configurations. We have verified that its effect on the signal is negligible. Consequently, all presented results were obtained with perfectly parallel moments. The orbital magnetic moments are an output of the calculation.

FDMNES automatically employs the magnetic space-group to define the equivalent and non-equivalent atoms, as well as the symmetry operations allowing the correspondence between them. Then, for the cross-section calculation, it works in real space using a cluster-based approach. All the atoms inside a sphere of a chosen radius $R$, centered onto the absorbing atom, are considered. We performed a series of calculations with increasing $R$, and found that convergence was achieved at $R = 6.24\,\text{\AA}$ (corresponding to 105 atoms in the cluster). All results shown in the paper are obtained with this radius. Due to the atomic displacements, symmetry is reduced, resulting in two non-equivalent Fe atoms in the unit cell. Therefore, two independent calculations must be performed for the two corresponding iron sites and their respective cluster of atoms. The final signal is given by:
\begin{equation}
A = \sum_{a}\sum_{s} S\!\left(A_a\right)
\label{eq:Tsum}
\end{equation}
where $A$ is the unit cell cross section, given in the Cartesian or spherical tensor form. $a$ indexes the two non-equivalent iron atoms, and $S$ stands for the symmetry operation giving the equivalent atoms tensor from the prototypical one $A_a$.

Calculations are relativistic and necessarily incorporate the spin-orbit interaction. We also account for the 0.032 eV splitting of the $1s^{\pm 1/2}$ core level, which arises from magnetically polarized exchange with the valence states. Although this splitting is small, to reproduce the experimentally observed extremely small signal ($\Delta I/I_{\mathrm{DC}}\approx 4\times10^{-6}$ at the pre-edge), an equivalent precision is required in all steps of the calculation. Indeed, the XAS signals recorded as a function of x-ray energy depend on the difference of energy between the photoelectron states and the core level states. Note that the eventual energy difference between the core states of the two non-equivalent sites is also considered, but this difference is negligible (approximately 0.001 eV), and likely has no discernible effect. We have used all the default FDMNES parameters for the convolution and the core all treatment (complete core all with a 1 electron screening in the first non-occupied, that is Fe $3d$ state).

\begin{figure}[tb]
  \centering
  \includegraphics[width=0.9\columnwidth]{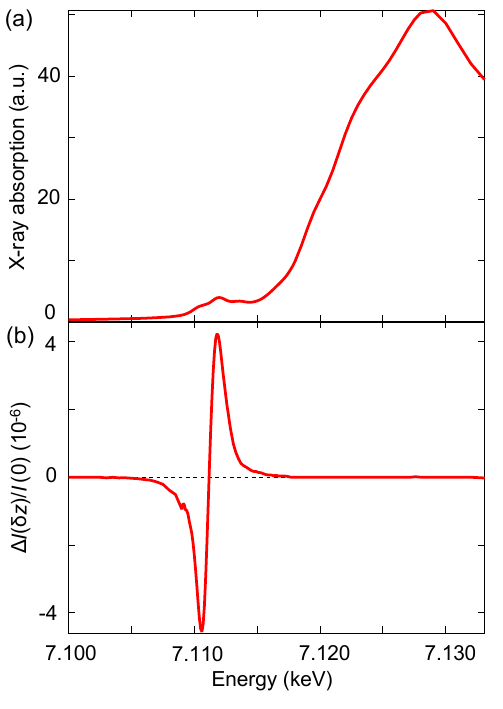}
  \caption{
   Numerical simulations of x-ray absorption spectrum and \textit{E}-induced XNDD spectrum for the $L_{\perp}+$ domain. (a) Simulated x-ray absorption spectrum without the Fe displacement. (b) Simulated spectrum of $\Delta I(\delta z)/I(0)$, corresponding to the wider energy range of the red curve shown in Fig. \ref{fig:XAS_domaindep}(c). Here, $\Delta I(\delta z)/I(0)$ denotes the normalized change in the transmitted x-ray intensity by introducing the Fe displacement $\delta z$ ($ = 5 \times 10^{-15}\,\mathrm{m}$) in the direction along the $c_h$ axis. 
  }
  \label{fig:sim_broad}
\end{figure}

Figure~\ref{fig:sim_broad} shows the simulated x-ray absorption spectrum (XAS) and the \textit{E}–induced XNDD spectrum over a wide energy range including the main edge. In the simulation of the \textit{E}-induced XNDD, the Fe displacement $\delta z$ is set at $5\times10^{-15}\,\mathrm{m}$. The simulated XAS agrees well with the experimental result [compare Figs.~\ref{fig:XAS_Edep}(a) and \ref{fig:sim_broad}(a)]. The \textit{E}-induced XNDD appears only in the pre-edge region.

\section{\textit{E}-induced XNDD measurements and simulations in the low-temperature phase}
\label{sec:lowT}

\begin{figure}[t]
  \centering
  \includegraphics[width=0.9\columnwidth]{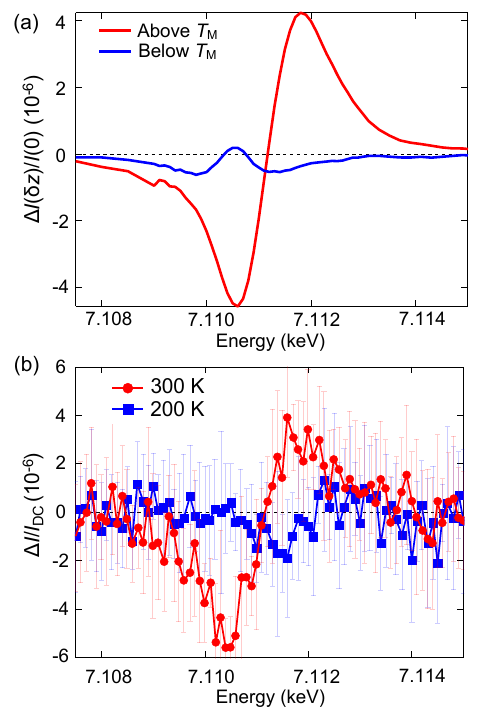}
  \caption{
   E-induced XNDD simulations and measurements in the low-$T$ phase. (a) Comparison of numerical simulations in the high-$T$ phase (red) and low-$T$ phase (blue). In the simulations, the Fe displacement is set to be identical in the two phases ($ \delta z= 5 \times 10^{-15}\,\mathrm{m}$). (b) Electric-field–modulated x-ray transmittance spectrum at 300 K ($L_{\perp}+$ domain, red) and 200 K (blue).  The spectra were obtained under an applied AC voltage with a peak amplitude of 200 V.
  }
  \label{fig:lowT}
\end{figure}

The magnetic structure of the low-$T$ phase below $T_\mathrm{M}$ also allows, by symmetry, the emergence of XNDD. However, numerical simulations show that the magnitude of the effect is one order of magnitude smaller than that in the high-$T$ phases [see Fig. \ref{fig:lowT}(a)]. At 200 K ($<T_\mathrm{M}$), we performed the same measurements as for the $L_{\perp}+$ state, with the x-ray beam linearly polarized along the $x$ axis and an AC voltage with a peak amplitude of 200 V applied. As shown in Fig. \ref{fig:lowT}(b), no meaningful signal was detected in the low-$T$ phase, indicating that \textit{E}-induced XNDD in the low-$T$ phase was below the detection limit of the present experimental setup. Although a small peak is discernible around 7.1115 keV, which is consistent with the numerical simulation, the overall signal amplitude is too weak to allow a reliable or detailed discussion of its spectral features.

\section{Spherical Multipole Description}
\label{sec:SphericalMultipole}

Here we show how the multipoles $I_{l,m}$ introduced in Eqs.~(\ref{eq:Mattertensor_1}) - (\ref{eq:Mattertensor_3}) are defined, taking Eq.~(\ref{eq:Mattertensor_2}) as an example. From the definition, Eq.~(\ref{eq:Mattertensor_2}) is written as
\begin{equation}
T_{xzx}-T_{yzy}
=2i\,\mathrm{Im}\sum_{n}\Big(
\langle xz\rangle_{0n}\langle x\rangle_{n0}
-\langle yz\rangle_{0n}\langle y\rangle_{n0}
\Big)\,
\delta_n.
\label{eq:Mattertensor_rewerite}
\end{equation}
For simplicity of notation, we shall remove the explicit dependence on $\delta_n$ in all expressions below. The terms $\langle xz \rangle_{0n}$ and $\langle yz \rangle_{0n}$ can be written as second-rank spherical tensors $T_{l,m}$ [e.g., $\langle xz \rangle_{0n} = \frac{1}{\sqrt{2}}(T_{2,-1}-T_{2,1})$ and $\langle yz \rangle_{0n} = \frac{i}{\sqrt{2}}(T_{2,-1}+T_{2,1})$ ], whereas $\langle x \rangle_{n0}$ and $\langle y \rangle_{n0}$ are first-rank spherical tensors [e.g., $\langle x \rangle_{n0} = \frac{1}{\sqrt{2}}(T_{1,-1}-T_{1,1})$ and $\langle y \rangle_{n0} = \frac{i}{\sqrt{2}}(T_{1,-1}+T_{1,1})$]. All tensors $T_{l,m}$ depend on $n$ (but for simplicity we have removed the dependence from the symbol). Therefore Eq.~(\ref{eq:Mattertensor_rewerite}) becomes $T_{xzx}-T_{yzy} =2i \mathrm{Im} \Sigma_n ( T_{2,-1} T_{1,-1}+ T_{2,1} T_{1,1})$. 
Then, by the addition theorem of spherical tensors, we can write a spherical-tensor product in terms of a single spherical tensor (call it $I_{l,m}$) as: $ T_{2,-1} T_{1,-1}= \sqrt{\frac{2}{3}}  I_{3,-2} + \frac{1}{\sqrt{3}}I_{2,-2}$ and $T_{2,1}T_{1,1}=\sqrt{\frac{2}{3}}  I_{3,2} - \frac{1}{\sqrt{3}}I_{2,2}$, where $\sqrt{\frac{2}{3}}$ and $\frac{1}{\sqrt{3}}$ are the Clebsch-Gordan coefficient. This defines the spherical tensors $I_{l,m}$.

\section{Orbital current interpretation of the magnetic toroidal octupole}
\label{sec:orbitalcurrent}

As shown in the main text, dominant contribution to the $E$-induced XNLD in hematite is the magnetic toroidal octupole  (specifically $I_{3,2} +I_{3,-2}$ in the $L_{\perp}+$ phase). Here we foucs on  the physical interpretation of the tensor $T_{xzx}-T_{yzy}$. Since the transition at the \textit{K}-edge takes place from the $1s$ state, where there is no spin-orbit coupling, the spin matrix elements can be factored out. Without spin contributions, the orbital part of the initial state becomes $\Psi_0\left(r,\ \theta,\ \varphi\right)=R_{1s}\left(r\right)Y_{0,0}\left(\theta,\varphi\right)=\sqrt{\frac{3}{4\pi}}R_{1s}(r)$. The intermediate states, $\Psi_n$, depend on the energy $\mathcal{E}$ and are written as $\Psi_\mathcal{E}\left(r,\ \theta,\ \varphi\right)=\sum_{l,m}{B_{l,m}\left(\mathcal{E}\right)R_{\mathcal{E},l}\left(r\right)Y_{l,m}\left(\theta,\varphi\right)}$. The angular part of the matrix elements in $T_{xzx}-T_{yzy}$ selects, via the Gaunt coefficients (i.e., integrals of triple spherical harmonics terms), exactly those coefficients $B_{l,m}\left(\mathcal{E}\right)$ that match the symmetry of the transition operator. This is a direct consequence of the initial $1s$ state whose spherical harmonics $Y_{0,0}\left(\theta,\varphi\right)$ is a constant. 

The matrix elements of Eq.~(\ref{eq:Mattertensor_rewerite}), rewritten from Eq.~(\ref{eq:Mattertensor_2}), are given by $\langle xz\rangle_{0n}=I_{R_2,\mathcal{E}}B_{2,xz}$, $\langle x\rangle _{n0}=I_{R_1,\mathcal{E}}B_{1,x}^*$, $\langle yz\rangle _{0n}=I_{R_2,\mathcal{E}}B_{2,yz}$, and $\langle y\rangle _{n0}=I_{R_1,\mathcal{E}}B_{1,y}^*$ where $I_{R_1,\ \mathcal{E}}$ and  $I_{R_2,\ \mathcal{E}}$ are energy depenedent radial integrals: $I_{R_2,\ \mathcal{E}}\ \propto\ \int_{0}^{\infty}{R_{1s}\left(r\right)r^4}R_{\mathcal{E},2}\left(r\right)$ and $I_{R_1,\ \mathcal{E}}\ \propto\ \int_{0}^{\infty}{R_{1s}\left(r\right)r^3}R_{\mathcal{E},1}\left(r\right)$. Therefore, $T_{xzx}-T_{yzy}$ is rewritten as
\begin{align}
T_{xzx}-T_{yzy}
&= I_{R_1,\mathcal{E}} I_{R_2,\mathcal{E}}
\Big[
\left(B_{2,xz}B_{1,x}^{\ast}-\mathrm{c.c.}\right)
\nonumber \\
&\qquad
-\left(B_{2,yz}B_{1,y}^{\ast}-\mathrm{c.c.}\right)
\Big]
\label{eq:Mattertensor_BR}.
\end{align}
This quantity in Eq.~(\ref{eq:Mattertensor_BR}) represents an orbital current in the intermediate state. Since the transition occurs at the \textit{K}-edge from the spinless $1s$ initial state (with negligible spin-orbit coupling), the current is purely orbital in origin.  The following example illustrates this interpretation. We consider the intermediate state $\Psi_\mathcal{E}\left(r,\ \theta,\ \varphi\right)$ and evaluate the quantum current on it. In general, the current is given by $\mathbf{J}=\frac{\hbar}{2m_ei}\left(\Psi^\ast\mathbf{\nabla}\Psi-\Psi\mathbf{\nabla}\Psi^\ast\right)$. Here $m_e$ is the electron mass. Applying this formula to the state $\Psi_\mathcal{E}\left(r,\ \theta,\ \varphi\right)$, we obtain $\mathbf{J}=j_r\mathbf{u}_r+j_\theta\mathbf{u}_\theta+j_\varphi\mathbf{u}_\varphi$ with
\begin{align}
j_r &= \frac{\hbar}{m_e}
\sum_{ll' m m'}
\mathrm{Im}\Big\{
B_{l' m'}^{\ast} B_{lm}
Y_{l' m'}^{\ast} Y_{lm}
R_{l'}\,\partial_r R_l(r)
\Big\},
\label{eq:jr}
\\
j_\theta &= \frac{\hbar}{m_er}
\sum_{ll' m m'}
\mathrm{Im}\Big\{
B_{l' m'}^{\ast} B_{lm}
R_{l'}R_l
Y_{l' m'}^{\ast}
\Big(
m\cot\theta\,Y_{lm}
\nonumber\\
&\qquad
+\sqrt{(l-m)(l+m+1)}e^{-i\varphi}Y_{l,m+1}
\Big)
\Big\},
\label{eq:jtheta}
\\
j_\varphi &= \frac{\hbar}{m_er\sin\theta}
\sum_{ll' m m'}
\mathrm{Im}\Big\{
B_{l' m'}^{\ast} B_{lm}
R_{l'}R_l
Y_{l' m'}^{\ast}Y_{lm}
(im)
\Big\}.
\label{eq:jphi}
\end{align}
By integrating Eqs.~(\ref{eq:jr})-(\ref{eq:jphi}) in space, we obtain the imaginary quantity $\Sigma_{ll^\prime m m^\prime}\mathrm{Im}\left(B_{l^\prime m^\prime}^\ast B_{lm}\right)$, which matches the form of Eq. (\ref{eq:Mattertensor_BR}) for some specific values of $l,\ l^\prime,\ m,\ m^\prime$. This implies that the value of the \textit{E}-induced XNLD represents the orbital current in the state $\Psi_\mathcal{E}$. 

\begin{figure}[tb]
  \centering
  \includegraphics[width=0.9\columnwidth]{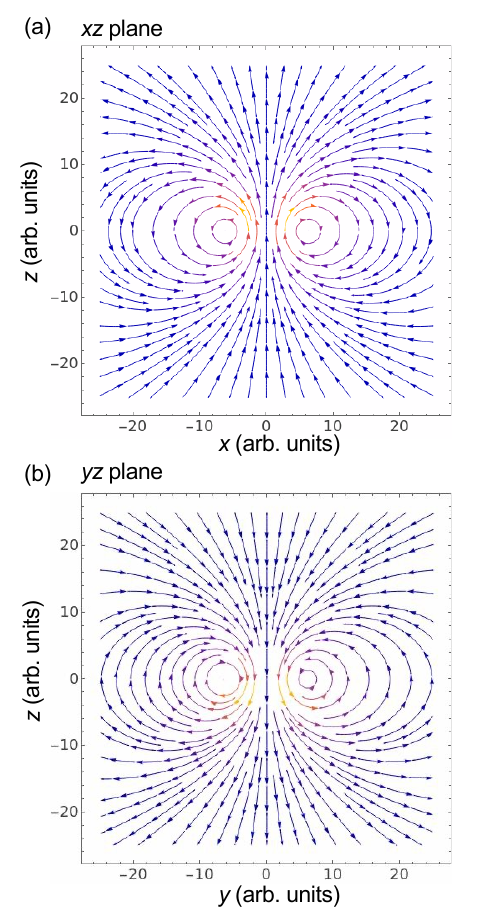}
  \caption{
   Streamline plots of orbital current of Eqs.~(\ref{eq:jr_final})-(\ref{eq:jphi_final}) in the $xz$ plane (a) and in the $yz$ plane (b). We remark the opposite behavior in the two planes: for positive $y$, the current flow rotates anticlockwise, whereas for positive $x$ it turns clockwise. Also for negative $y$, the current flow rotates clockwise, whereas for negative $x$ it turns anticlockwise. The units of horizontal and vertical axes are arbitrary units.
  }
  \label{fig:current}
\end{figure}

To understand what kind of orbital current in Eq.~(\ref{eq:Mattertensor_BR}) represents, we consider a wavefunction $\Psi_\mathcal{E}$ with only $p_x, p_y, d_{xz}, d_{yz}$ orbitals, so as to reproduce the features of the result in Eq.~(\ref{eq:Mattertensor_BR}). $\Psi_\mathcal{E}\left(r,\ \theta,\ \varphi\right)$ can be rewritten using the real spherical harmonics $Y_{p_x},\ Y_{p_y},\ Y_{d_{xz}},\ \mathrm{and}\ Y_{d_{yz}}$, and consequently Eqs.~(\ref{eq:jr})-(\ref{eq:jphi}) are rewritten as
\begin{equation}
\begin{aligned}
j_r &= \frac{2\hbar}{m_e}\left(R_pR_d' - R_dR_p'\right)
\frac{3\sqrt{5}}{4\pi}\sin^2\theta\cos\theta
\\
&\times
\Big[
\mathrm{Im}\!\left(a_x^\ast b_{xz}\right)\cos^2\varphi
+\mathrm{Im}\!\left(a_y^\ast b_{yz}\right)\sin^2\varphi
\\
&\qquad
+\mathrm{Im}\!\left(a_x^\ast b_{yz}+a_y^\ast b_{xz}\right)
\sin\varphi\cos\varphi
\Big]
\end{aligned}
\label{eq:jr_final}
\end{equation}
\begin{equation}
\begin{aligned}
j_\theta &= \frac{2\hbar}{m_er}R_pR_d
\frac{-3\sqrt{5}}{4\pi}\sin^3\theta
\\
&\times
\Big[
\mathrm{Im}\!\left(a_x^\ast b_{xz}\right)\cos^2\varphi
+\mathrm{Im}\!\left(a_y^\ast b_{yz}\right)\sin^2\varphi
\\
&\qquad
+\mathrm{Im}\!\left(a_x^\ast b_{yz}+a_y^\ast b_{xz}\right)
\sin\varphi\cos\varphi
\Big]
\end{aligned}
\label{eq:jtheta_final}
\end{equation}
\begin{equation}
\begin{aligned}
j_\varphi &= j_\varphi^{\mathrm{odd}}+j_\varphi^{\mathrm{even}}
\\
&= \frac{2\hbar}{m_er}\frac{3}{4\pi}\sin\theta
\Big\{
R_pR_d\sqrt{5}\cos\theta
\,\mathrm{Im}\!\left(a_x^\ast b_{yz}-a_y^\ast b_{xz}\right)
\\
&\qquad
+\left[
R_p^2\,\mathrm{Im}\!\left(a_x^\ast a_y\right)
+5\cos\theta\,R_d^2\,\mathrm{Im}\!\left(b_{xz}^\ast b_{yz}\right)
\right]
\Big\}
\end{aligned}
\label{eq:jphi_final}
\end{equation}
Here, $j_\varphi^{\mathrm{even}}$ in Eq.~(\ref{eq:jphi_final}) is the parity-even part of the current, which therefore, does not contribute to the XNDD. In the $L_{\perp}$ state with application of $E$, only the coefficients with an even number of $x$ labels in Eqs.~(\ref{eq:jr_final})-(\ref{eq:jphi_final}) survive (see also Sec.~\ref{Sec:tensorAna}).

\begin{figure}[t]
  \centering
  \includegraphics[width=0.8\columnwidth]{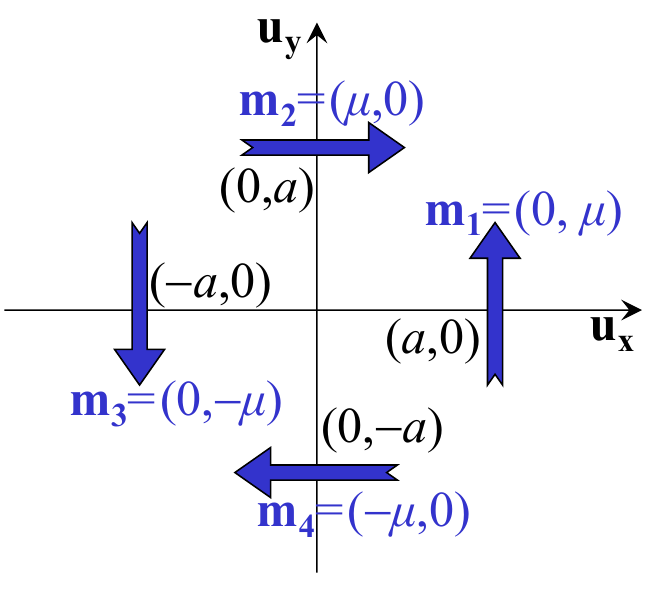}
  \caption{
   Multipole pattern corresponding to the orbital current shown in Eqs.~(\ref{eq:jr_final})-(\ref{eq:jphi_final}) and Fig.~\ref{fig:current}. The set of magnetic dipoles at the given positions corresponds to a linear combination of the $M_{xy}$ component of the magnetic quadrupole and the $O_{x^2-y^2,\ z}$ component of the magnetic toroidal octupole.
  }
  \label{fig:Multipole}
\end{figure}

A simple choice to implement the symmetry of Eqs.~(\ref{eq:jr_final})-(\ref{eq:jphi_final}) is to take $a_x=b_{yz}=1$ and $a_y=b_{xz}=i$. Then we obtain $j_\varphi^{\mathrm{odd}}=0$, $j_\theta\propto\sin^3{\theta}\cos{(2\varphi)}$ and $j_r\propto\sin^2{\theta}\cos{\theta}\cos{\left(2\varphi\right)}$. The streamline plots of the current $\mathbf{J}$ for this choice of coefficients in the $xy$ and $yz$ planes are shown in Fig.~\ref{fig:current}. Using Ampere’s theorem (a closed loop current is equivalent to a magnetic dipole), this figure suggests the presence of magnetic dipoles at the positions schematically represented, in Fig.~\ref{fig:Multipole} in the $xy$ plane.

Finally, we evaluate the multipolar components associated to Fig.~\ref{fig:Multipole}. The magnetic toroidal dipole ($t_k$), the magnetic quadrupole ($M_{ij}$), and the magnetic toroidal octupole ($O_{ij, k}$) are defined as
\begin{align}
t_k &= \epsilon_{ijk} t_{ij}
= \epsilon_{ijk}\frac{1}{2}
\sum_\alpha
\left(
r_i^{(\alpha)} m_j^{(\alpha)}
- m_i^{(\alpha)} r_j^{(\alpha)}
\right)
\label{eq:tk}
\\
M_{ij} &= \frac{1}{2}
\sum_\alpha
\Bigg(
r_i^{(\alpha)} m_j^{(\alpha)}
+ m_i^{(\alpha)} r_j^{(\alpha)}
\nonumber\\
&\qquad
- \frac{2}{3}\delta_{ij}
\sum_l m_l^{(\alpha)} r_l^{(\alpha)}
\Bigg)
\label{eq:Mij}
\\
O_{ij,k} &=
\sum_\alpha
\left(
r_i^{(\alpha)} r_j^{(\alpha)}
- \frac{1}{3}\delta_{ij}
\sum_l r_l^{(\alpha)} r_l^{(\alpha)}
\right)
t_k^{(\alpha)}
\label{eq:Oijk}
\end{align}
where $\alpha$ labels the positions of the magnetic dipole $m_i$, and $\epsilon_{ijk}$ is the Levi-Civita symbol. Using Eqs.~(\ref{eq:tk})-(\ref{eq:Oijk}), the magnetic multipoles of Fig.~\ref{fig:Multipole} are evaluated as $O_{x^2-y^2,\ z}=2\mu a^3$, $M_{xy}=2\mu a$, and all other multipole components are identically zero. These are the Cartesian components of the multipoles detected in the experiment: $O_{x^2-y^2,\ z}\rightarrow I_{3,2}+I_{3,-2}$ and $M_{xy}\rightarrow I_{2,2}-I_{2,-2}$. 

\nocite{*}
\clearpage
\bibliographystyle{apsrev4-2}
\bibliography{paperpile}

\end{document}